\documentclass{resonance}

\usepackage{physics}
\begin{document}

%%paper title
%%For line breaks, \\ can be used within title 
\title{A Concise History of the Black-body Radiation Problem}
%\secondTitle{Second Title here}
\author{Himanshu Mavani and Navinder Singh}

\maketitle
%%\authorIntro is used to place the author's photo and an introduction about the author
%%photo goes into the includegraphics with width=2cm 
%%and a "\\" dividing the text and photo
%%the intro text box is drawn automatically
%%place \authorIntro  just before abstract
\authorIntro{\includegraphics[width=2cm, height=2.5cm]{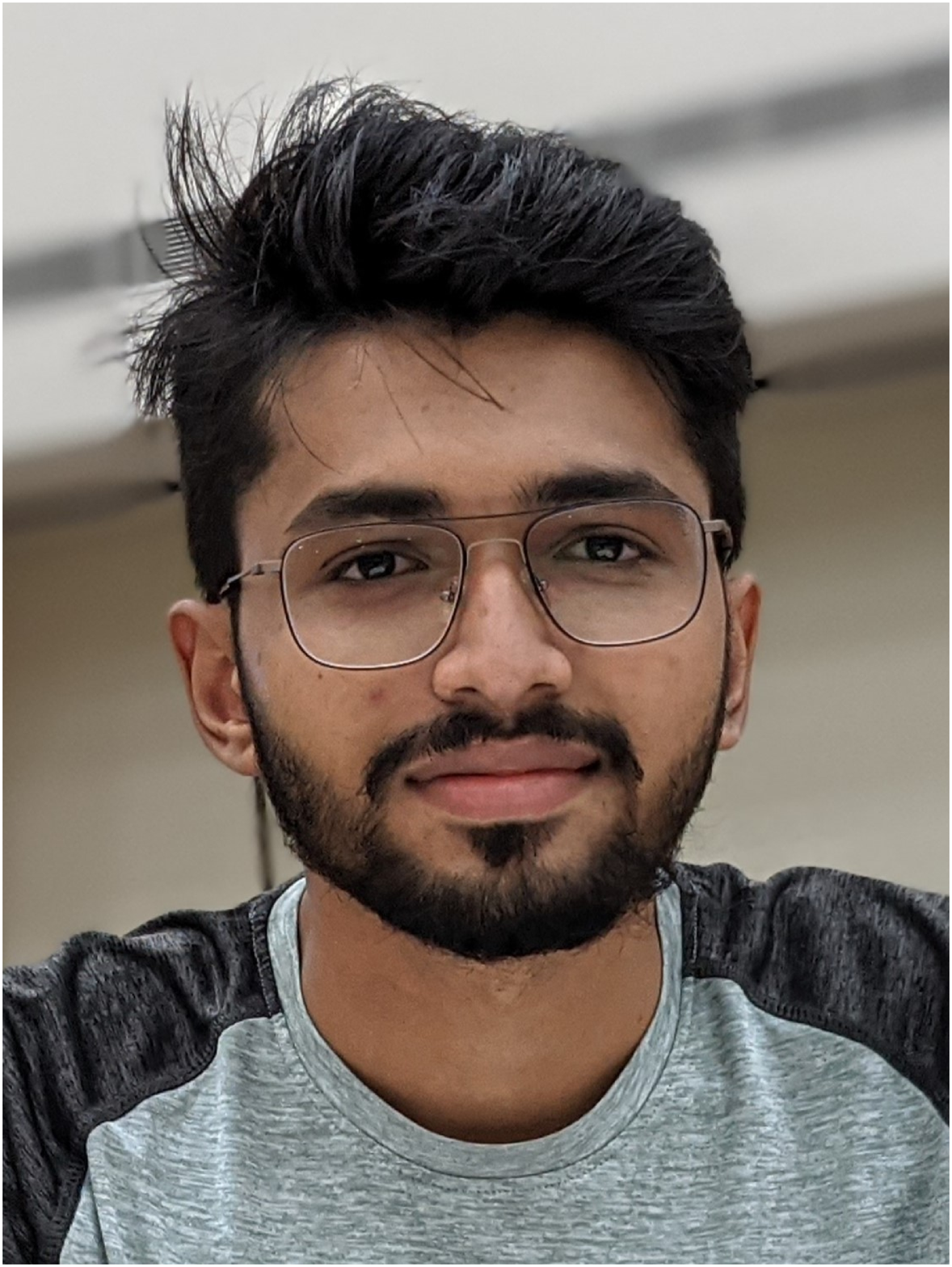}\\Himashu Mavani is presently a Integrated M.Sc. student in School of Physical Sciences at National Institute of Science Education and Research, Bhubaneshwar. He is interested in theoretical condensed matter physics.\\
\vspace{0.5cm}
\includegraphics[width=2cm, height=2.5cm]{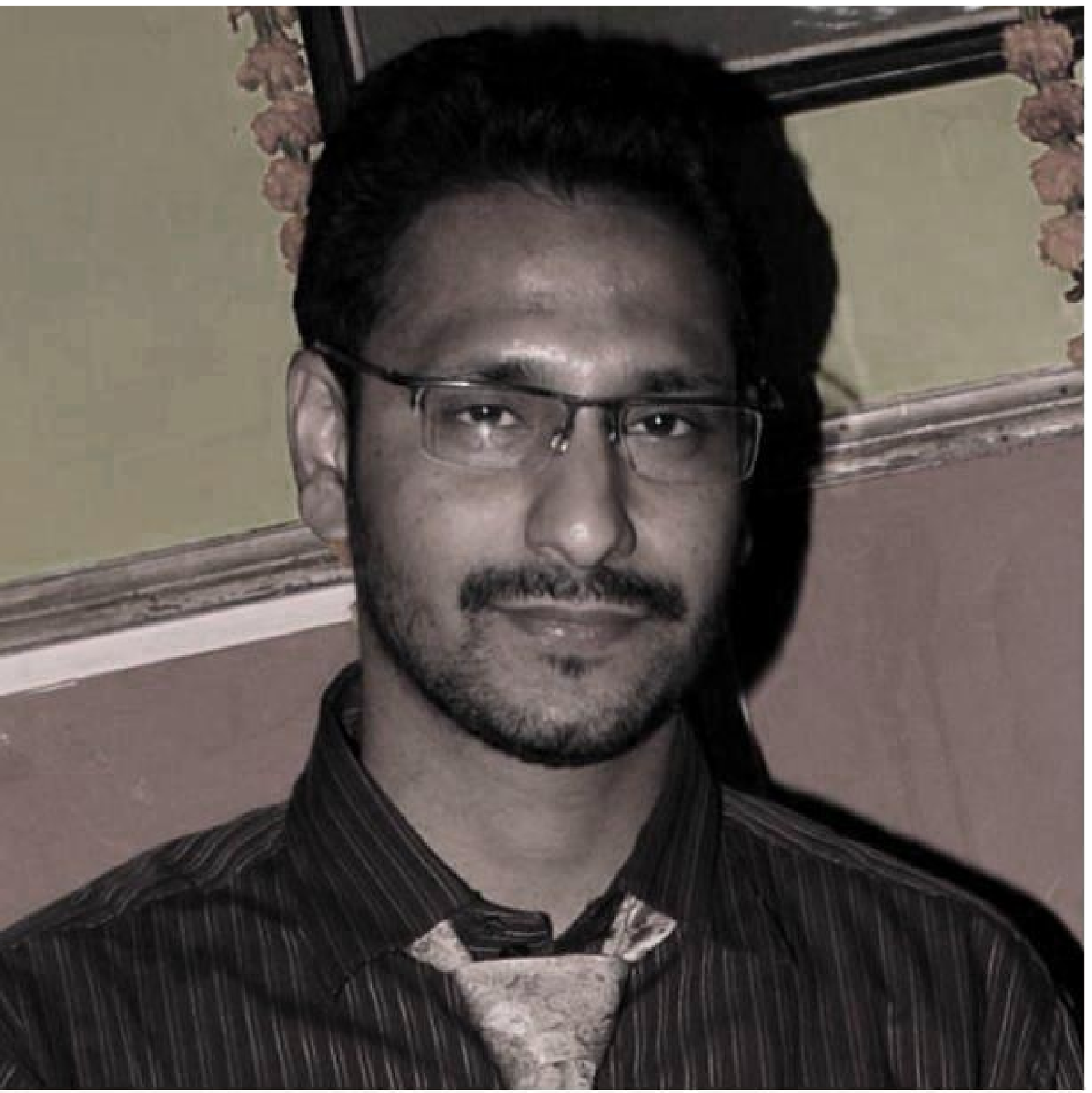}\\Navinder Singh works in the Physical Research Laboratory, Ahmedabad. His research interests are in non-equilibrium statistical mechanics and quantum dissipative systems .}
%%abstract
\begin{abstract}
The way the topic of black-body radiation is presented in standard textbooks (i.e. from Rayleigh-Jeans to Max Planck) does not follow the actual historical timeline of the understanding of the black-body radiation problem. Authors believe that a presentation which follows an actual timeline of the ideas (although not a logical presentation of the field) would be of interest not only from the history of science perspective but also from a pedagogical perspective. Therefore, we attempt a concise history of this very interesting field of science.
\end{abstract}

%%include \monthyear{month year} for month and year of publication in the footer
\monthyear{August 2022}
%%use \artNature for the running head information
\artNature{GENERAL  ARTICLE}

\section*{Introduction}
Generally, we understand Planck's distribution function for the black-body radiation as a modification of Rayleigh-Jeans law using the canonical ensemble of quantized energy. However, as the foundation theory of quantum mechanics, it is worth understanding the actual development of the ideas starting from the pioneering work of Gustav Kirchhoff in this field. Kirchhoff's theoretical analysis gave scientists a way to characterize the black-body spectrum. Josef Stefan's empirical deduction of the $T^4$ law initiated further theoretical investigations. Ludwig Boltzmann's thermodynamic derivation of Josef Stefan's law verifies the idea of radiation pressure. Next, the fascinating derivation of Wilhelm Wien's scaling law (but lesser known) is discussed, which led Max Planck to conclude that the entropy of an oscillator is a function of the ratio of the average energy and frequency of an oscillator. Next, we review Planck's original derivation of his distribution function to show the historical importance of Kirchhoff's law, Stefan-Boltzmann's $T^4$ law, and Wien's scaling law. Finally, the classical distribution function of Lord Rayleigh and James Jeans is discussed. In the following paragraphs, we briefly sketch this historical development. In the following sections, a more detailed analysis is given.\\

The theory of radiative heat exchange was first initiated by Genevan physicist Pierre Prevost in 1791 \cite{Maxwell_Book, chang_2002}. He defined the thermal equilibrium in the context of radiative heat transfer and explained that each body radiates and receives heat independently of the presence of the other bodies. The thermal radiation is in the infrared frequency range when the temperature of the body is equal to room temperature, so we can not see that. As we increase the temperature, the body starts gloving from red to white in colour. In 1830, Leopoldo Nobili and Macedonio Melloni made a thermopile device, which converts thermal energy into electrical current. In 1831, they made the first radiometer using thermopile and galvanometer \cite{Leopoldo_1831}. Their device showed that the amount of radiation emitted from different surfaces at the same temperature is not the same. In 1847 John William Draper observed that $525^0C$ is an average temperature where emission radiation becomes visible, and it is known as $\textit{Draper point}$ \cite{draper_1847}. In 1858, Scottish physicist Balfour Stewart experimentally measured the radiation using thermopile and compared thermal emission and absorptive power of different materials with lamp-black \cite{stewart_1858, siegel_1976, robitaille_2008}. Stewart wrote, ``Lamp-black which absorbs all the rays that fall upon it, therefore, possesses the greatest possible absorbing power and posses the greatest possible radiating power".\\

The notion of a black-body was first defined in a concrete way by Gustav Kirchhoff: A body which reflects no light at all nor allows light to pass. No ideal black-body exists in nature. Lamp-black and Platinum-black are good approximations. Kirchhoff theoretically explained that the emission spectrum (energy density) is independent of the shape, size and material of the black-body, and it is the only function of radiation wavelength and temperature. Thus, if two different black bodies are in equilibrium, then their radiation field is identical.\\

John Draper, John Tyndall and many other physicists have studied the temperature dependence of energy density of thermal radiation. Draper also plotted the data on total energy emission vs temperature in 1847 \cite{draper_1847}.\rightHighlight{Draper's data was on the Fahrenheit scale. The absolute thermometric scale (Kelvin scale) was invented by William Lord Kelvin in 1848.} Josef Stefan empirically described the dependence of temperature on the total energy emitted by a black-body. Then five years later, Ludwing Boltzamnn theoretically proved Stefan's law using purely thermodynamic arguments. This law is known as Stefan-Boltzmann $T^4$ law.\\

Wilhelm Wien is known for finding the relation between temperature and wavelength, where the energy density is maximum. Apart from this Wien's displacement law, he was the first who gave the simplified form of the Kirchhoff universal function using rigorous thermodynamical arguments in 1893. He also gave the empirical energy distribution for the black-body in 1896, but it turned out to be just an approximation.\\

In June 1900, Lord Rayleigh considered black-body radiation as electromagnetic standing wave vibrations in cavity enclosure, and he used the equipartition theorem of statistical mechanics. He found that the energy distribution function of black-body radiation is proportional to the $T\lambda^{-4}$. Rayleigh and English physicist James Jeans found a complete form of the energy distribution function five years later, but it disagreed with the experimental observations.\\

German physicist Max Planck has been working on the black-body problem for more than five years. He assumed that cavity walls are made of a collection of oscillators of electric dipoles. He published his exact solution for the black-body problem in two papers (October 1900 and December 1900), where he introduced the idea of energy quantization. Planck's implication on the quantum of energy and historical debates on it is discussed by Thomas S. Kuhn \cite{kuhn_1987}, Luis J. Boya \cite{boya_2004}, M. J. Klein \cite{klein_1961}, Allan Needell \cite{needell_980}, Olivier Darrigol \cite{darrigol_1988}, and Clayton A. Gearhart \cite{gearhart}.\\

In the subsequent section, we consider the contributions of the leading investigators in this field one by one.

\section{Enter Gustav Kirchhoff}
\begin{figure}[h]
	\caption{Gustav Kirchhoff \textit{(1824-1887) [Photo: Wikipedia Commons]}}\label{Kirchhoff}
	\vskip -12pt
	\centering
	\includegraphics[width=4.2cm, height=5.5cm]{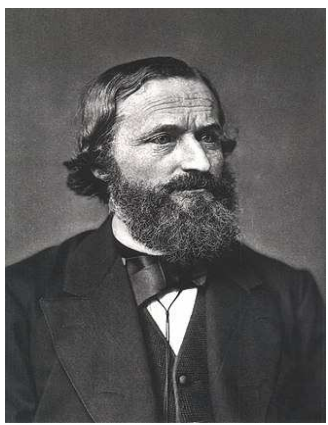}
\end{figure}

In 1859, German physicist Gustav Robert Kirchhoff gave the first theoretical argument on the back-body radiation problem.\leftHighlight{The Bunsen–Kirchhoff Award in the field of analytical spectroscopy is named after Robert Bunsen and Gustav Kirchhoff.}Along with Robert Bunsen, Kirchhoff studied the coincidence of the emission lines of sodium with particular absorption lines in the solar spectrum using a prism. From his experiments, he considered that at thermal equilibrium, the ratio of emissive power to absorptive power is the same for all bodies and asked the question: does this the case for each wavelength separately? He concludes that for any thermally radiating body, the ratio of emissive power ${e}$ and the absorptive power ${a}$ is a universal function that only depends upon the wavelength and temperature \cite{kirchhoff_1859, Kirchhoff_1860}. \\

\begin{figure}[h]
	\caption{Kirchoff's setup: Plate $C$ of emissive and absorbing power $E$ and $A$ respectively. Plate $c$ of emissive and absorbing power $e$ and $a$ respectively.}
	\centering
	\includegraphics[width=4cm, height=5cm]{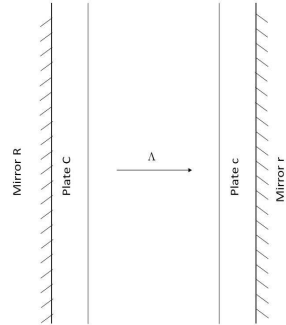}
	\label{setup1}
\end{figure}
Kirchhoff considered two thin plates ${C}$ and ${c}$ of emissive power ${E}$ and ${e}$ and absorptive power ${A}$ and ${a}$ respectively \cite{kirchhoff_1859, siegel_1976}. 
One of the plate surfaces is a perfectly reflecting mirror, and the plates are arranged parallel to each other, as shown in Figure $(2)$. Plate ${c}$ is made of special material that only absorbs and emits the wavelength radiation $\Lambda$. All other wavelengths pass through. Kirchhoff considered a quantity of radiation $E$ of wavelength $\Lambda$ emitted by ${C}$. When it comes to plate $c$, it absorbs $aE$ radiation and throws $(1-a)E$ radiation back. When this $(1-a)E$ radiation reach to plate $C$, it absorbs $A(1-a)E$ radiation and throws $(1-A)(1-a)E$ then again plate $c$ absorbs $a(1-A)(1-a)E$ radiation and throws $(1-a)(1-A)(1-a)E$ radiation and so on. This infinite series of bounces back and forth of radiation between two mirrors gives the total absorption by plate $c$ as
\begin{equation}
	\frac{aE}{1-(1-A)(1-a)}\ .
\end{equation}
Similarly, if the plate $c$ emits the quantity of radiation $e$ of wavelength $\Lambda$, then the plate $c$ absorbs radiation of quantity 
\begin{equation}
	\frac{a(1-A)e}{1-(1-A)(1-a)}\ .
\end{equation}
So at thermal equilibrium, total absorption by the plate $c$ should be equal to $e$.
\begin{equation}
	e = \frac{aE}{1-(1-A)(1-a)} + \frac{a(1-A)e}{1-(1-A)(1-a)}
\end{equation}
By rearranging the terms, we get
\begin{equation}
	\frac{e}{a} = \frac{E}{A}
\end{equation}
and this is true for any plate $C'$ with emissive power ${E'}$ and absorptive power ${A'}$. 
\begin{equation}
	\frac{e}{a} = \frac{E'}{A'}
\end{equation}
This shows that for the same temperature and wavelength, the ratio of emissive power ${e}$ and the absorptive power ${a}$ is the same for all bodies. i.e. it is  universal function $f(\lambda, T)$. 
\begin{equation}
	\frac{e}{a} = f(\lambda, T)\\
	\label{K1}
\end{equation}

After one month, in Jan-1860, Kirchhoff published a second paper where he proved this theorem in general context\cite{Kirchhoff_1860, kelly_1965, siegel_1976, schirrmacher_2003}. In the second paper, he introduced the term black-body, which absorbs all of the radiation falling upon it, so its absorptive power is one. Then the emission power of the black-body is a universal function of wavelength and temperature. Collecting contributions from the whole spectrum (integrating over all wavelengths), we can see that the total energy emitted per unit volume by the black-body is the only function of its temperature, which is given by the Stefan-Boltzmann $T^4$ law.

\section{Enter Josef Stefan}
\begin{figure}[h]
	\caption{Josef Stefan \textit{(1835-1893) [Photo: Wikipedia Commons]}}
	\centering
	\includegraphics[width=4.2cm, height=5.5cm]{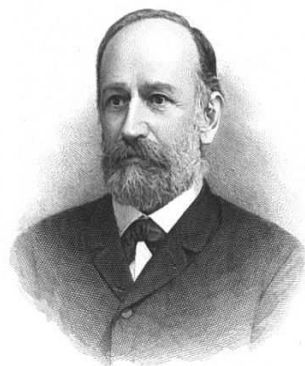}
	\label{Stefan}
\end{figure}
In 1864, Irish physicist John Tyndall did measurements of the infrared emission by the platinum filament, and the corresponding colour of filament \cite{tyndall_1865}. \rightHighlight{John Tyndall was a pioneering experimental physicist in the field of atmospheric physics. He is known for discovering the greenhouse effect and the Tyndall effect.}In 1879, Austrian physicist Josef Stefan gave an empirical law of temperature dependence of the black-body radiation based on Tyndall and various other experimental data. In Tyndall's measurements, Stefan noticed that by increasing the temperature of platinum filament from $798 K$ to $1473 K$ ($1.845$ times), radiation was increased by $11.7$ times (approximate to $(1.845)^4$). So he empirically stated that the energy emitted by a black-body per unit area per second is proportional to the fourth power of the absolute temperature \cite{stefan_1879}.

\section{Enter Ludwig Boltzmann}
\begin{figure}[h]
	\caption{Ludwig Boltzmann \textit{(1844-1906) [Photo: Wikipedia Commons]}}
	\centering
	\includegraphics[width=4.2cm, height=5.5cm]{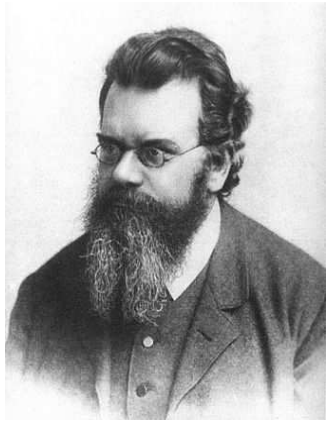}
	\label{Boltzmann}
\end{figure}

In 1884, Stefan's formal doctoral student Ludwig Boltzmann published a thermodynamic derivation of Stephen's empirical law. Using the idea of Adolfo Bartoli on radiation pressure \cite{bartoli_1884}, Boltzmann considered radiation particles as the ideal gas in the piston-cylinder system. \rightHighlight{The concept of radiation pressure on a solid body was experimentally verified by Pyotr Lebedev in 1899 \cite{masalov_2019}.}\\
The pressure exerted on the piston by the radiation particles when the piston moves slowly is
\begin{equation}
	P = \frac{1}{3}\rho_T
\end{equation}
where $\rho_T$ is the total energy density (Appendix A).\\

Putting total internal energy $U = \rho_T V$ and radiation pressure into the fundamental thermodynamics relation $dU = TdS - PdV$, we get
\begin{equation}
	TdS = \frac{4}{3}\rho_TdV + V\dv{\rho_T}{T}dT 
	\label{B1}
\end{equation}
We know $ S \equiv S(T,V)$ so
\begin{equation}
	dS = \pdv{S}{T}dT + \pdv{S}{V}dV
	\label{B2}
\end{equation}
By comparing equation (\ref{B1}) and (\ref{B2}), we get 
\begin{equation}
	\pdv{S}{T} = \frac{V}{T}\dv{\rho_T}{T}, \quad \pdv{S}{V} = \frac{4}{3}\frac{\rho_T}{T}\ .
\end{equation}
So we have
\begin{equation}
	\pdv{S}{V}{T} = \frac{1}{T}\dv{\rho_T}{T}, \quad \pdv{S}{T}{V} = -\frac{4}{3}\frac{\rho_T}{T^2} + \frac{4}{3}\frac{1}{T}\dv{\rho_T}{T}\ .
\end{equation}
By symmetry of partial derivative, equating above terms, we get
\begin{equation}
	\dv{\rho_T}{T} = 4\frac{\rho_T}{T}\ .
\end{equation}
Integrating the above equation, we get
\begin{equation}
	\rho_T = a T^4\ .
\end{equation}
Using Lambert's relation $\rho_T = \frac{4}{c}E$ (where $E$ is the total Emissive power over the hemisphere) we get the well-known Stefan-Boltzmann law $E = \sigma T^4$. The constant $\sigma = \frac{ac}{4}$ is known as Stefan-Boltzmann constant. The exact form of a constant can not be determined from purely classical mechanics.\\

\section{Enter Wilhelm Wien}
\begin{figure}[h]
	\caption{Wilhelm Wien \textit{(1864-1928) [Photo: Wikipedia Commons]}}
	\centering
	\includegraphics[width=4.2cm, height=5.5cm]{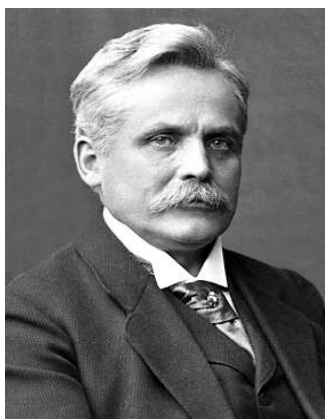}
	\label{Wien}
\end{figure}
The German physicist Wilhelm Franz Wien's work has been the most important work in studying black-body radiation. In 1893, Wien used thermodynamical arguments and the Doppler effect and derived important relation between the energy density of radiation with frequency and temperature.
\begin{equation}
	\rho(\nu) = \nu^3f(\nu/T)
\end{equation}
This should be called Wien's scaling law. Using this, he also derived that the product of the wavelength where the emission spectrum of black-body has maximum and temperature is constant, known as ``Wein's displacement law". \\

Now we present a concise derivation of Wien's scaling law which is hard to find in the books.\mfnote{Only Max Born discussed this briefly in his book "Atomic Physics".} As before, suppose radiation is enclosed in a piston-cylinder system and is in thermodynamical equilibrium at temperature T. \\

\begin{figure}[h]
	\caption{The radiation inside the piston-cylinder system. $S$ denotes a radiation source, $O$ denotes an observer, and $u$ is the velocity of the piston.}
	\centering
	\includegraphics[width=5cm, height=3cm]{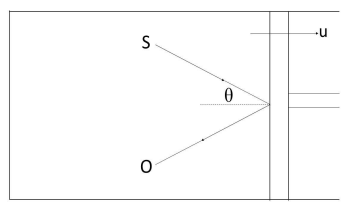}
	\label{setup2}
\end{figure}
Suppose that the inner surface of the piston is perfectly reflected. Radiation exerts radiation pressure on the piston. A piston is allowed to move out quasi-statically, and this process is made adiabatic, so no external heat is allowed to enter into the cylinder. Suppose at time $t=0$ a radiation beam coming from point $S$ is incident on a mirror at angle $\theta$ with the normal to the piston. The observer at $O$ receives the piston reflected beam from the moving mirror. Let $u$ be a velocity of the piston. Now, the Doppler effect with velocity $ucos\theta$ since it is relevant to observer $O$ gives the frequency observed by $O$.
\begin{equation}
	\nu' = \nu - \frac{2u}{c}\nu cos\theta
	\label{WW1}
\end{equation}
Let us denote the $I(\nu)d\nu$ as intensity of radiation whose frequency is in range $\nu$ and $\nu + d\nu$. Then the energy in the range $\nu$ and $\nu + d\nu$ falling on the patch of area $dA$ on the piston mirror in time $dt$ is
\begin{equation}
	I(\nu)d\nu dA_{\perp}dt = I(\nu)cos\theta dAdtd\nu
\end{equation}
If $I(\nu')d\nu'$ is the intensity of the reflected radiation whose frequency lies in $\nu'$ to $\nu' + d\nu'$, the reflected energy from patch in time $dt$ is
\begin{equation}
	I'(\nu')d\nu'\ dA_{\perp}dt = I'(\nu')cos\theta dAdtd\nu'
\end{equation}
then loss of the radiation energy will be 
\begin{equation}
	-(I(\nu)d\nu - I'(\nu')d\nu')cos\theta dAdt\ .
	\label{w1}
\end{equation}
This loss appears in the form of work done on the piston by radiation. The pressure exerted by radiation on the piston is 
\begin{equation}
	dP = 2\rho(\nu)\ cos^2\theta\ d\nu\ .
\end{equation}
So, the work done will be
\begin{equation}
	2\rho(\nu)\ cos^2\theta\ d\nu\ dA\ udt\ .
\end{equation}
Equating this with the equation (\ref{w1}) and using $I(\nu) = \rho(\nu)c$ we get
\begin{equation}
	I'(\nu')d\nu' = I(\nu)\left(1 + \frac{2ucos\theta}{c} \right)d\nu
\end{equation}
Again using the $I(\nu) = \rho(\nu)c$ and integrating both side, we get 
\begin{equation}
	\rho' = \rho\cdot\left(1 + \frac{2ucos\theta}{c} \right)
\end{equation}
where $\rho'$ and $\rho$ denote the integrated energy density. So, after reflection, energy density is increased by a factor $\left(1 + \frac{2ucos\theta}{c} \right)$.\\

To calculate the total energy change in time $dt$ of radiation whose frequency is in the range $\nu$ to $\nu + d\nu$, we need to take an angular average of the difference in final and initial state energy, which means
\begin{equation}
	\Delta E = \int_{\phi = 0}^{2\pi}\int_{\theta = 0}^{\pi/2} c dt\ dA\ cos\theta\ \frac{(sin\theta d\theta d\phi)}{4\pi} \left(\rho'(\nu')d\nu' - \rho(\nu)d\nu \right)
	\label{wa}
\end{equation}
All the radiation whose frequency is in the range $\nu$ to $\nu + d\nu$, after reflection that will be in range $\nu'$ to $\nu' + d\nu'$, where $\nu'$ is given by equation (\ref{WW1}). So the term $\rho'(\nu')d\nu'$ is replaced by
\begin{equation}
	\left(1 + \frac{2ucos\theta}{c} \right)\rho\left(\nu - \frac{2ucos\theta}{c}\nu\right)\ \left(1 - \frac{2ucos\theta}{c}\right)d\nu\ .
	\label{w5}
\end{equation}
Writing Taylor expansion,
\begin{equation}
	\rho\left(\nu - \frac{2ucos\theta}{c}\nu\right) = \rho(\nu) - \frac{2ucos\theta}{c}\nu\pdv{\rho(\nu)}{\nu} - \cdots
\end{equation}
and neglecting terms of $\frac{u^2}{c^2}$ and higher in equation (\ref{w5}), and putting into equation (\ref{wa}), we get
\begin{equation}
	\Delta E = \frac{cdt\ dA}{2}\int_{\theta=0}^{\pi/2} d\theta sin\theta cos\theta \left(-2\frac{u cos\theta}{c}\nu\pdv{\rho(\nu)}{\nu} \right)
\end{equation}
Now using the $dV = u\ dt\ dA$ and $\int_{0}^{\pi/2} d\theta sin\theta cos^2\theta =\frac{1}{3} $, we get
\begin{equation}
	\Delta E = -dV\frac{\nu}{3}\pdv{\rho(\nu)}{\nu}d\nu
\end{equation}
The negative sign shows that there is a net reduction in energy which appears as work done by the piston.  The reduction of the energy for the frequency range $\nu$ to $\nu + d\nu$ can be written as $-d(\rho(\nu)V)d\nu$. So we have
\begin{equation}
	d(\rho(\nu)V)d\nu = dV\frac{\nu}{3}\pdv{\rho(\nu)}{\nu}d\nu
\end{equation}
since the process of moving the piston out by a very small amount $vdt$ is adiabatic, we can take $\rho(\nu)$ as the only function of volume since the energy density reduces with an increase in volume.
\begin{equation}
	\rho(\nu) + V\pdv{\rho(\nu)}{V} = \frac{\nu}{3}\pdv{\rho(\nu)}{\nu}
\end{equation}
It easy to check that function $\rho(\nu) = \nu^3\phi(\nu^3V)$ is a solution of this equation.\\
Now, using the equation (\ref{B1}) for adiabatic process ($ds = 0$) we get
\begin{equation}
	0 = \frac{V}{T}\pdv{\rho}{T}dT + \frac{4}{3}\frac{\rho}{T}dV
\end{equation}
Now, using the Stefan-Boltzmann law, we get
\begin{equation}
	0 = \frac{V}{T}(4a T^3)dT + \frac{4}{3}\frac{a T^4}{T}dV
\end{equation}
\begin{equation}
	\text{So,} \quad \frac{dT}{T} = -\frac{1}{3}\frac{dV}{V} 
\end{equation}
which gives the solution $VT^3 =$ constant.

%%%%%%%

So, we have 
\begin{equation}
	\rho(\nu) = \nu^3\phi\left((const)\frac{\nu^3}{T^3} \right) \equiv \nu^3f\left(\frac{\nu}{T} \right)
	\label{w2.0}
\end{equation}
Using $\rho(\nu)d\nu = \rho(\lambda)d\lambda$ and $\lambda = c/\nu$, we get
\begin{equation}
	\rho(\lambda, T) = \frac{c^4}{\lambda^5}f\left(\frac{c}{T\lambda} \right)
	\label{w2}
\end{equation}
Thus Wien's scaling law simplifies Kirchhoff's universal function. This suggests that a plot of $\lambda^5\rho(\lambda, T)$ vs $\lambda T$ gives a single curve regardless of the wavelength of radiation and temperature of the black-body. In other words, knowing a spectrum for a single temperature, a spectrum can be found for any other temperature. Later, other equivalent forms were given by Lord Rayleigh, and Max Ferdinand Thiesen \cite{rayleigh_1889}.\\

Let $\rho(\lambda)$ is maximum at $\lambda_m$. Then derivative of $\rho(\lambda)$ with respect to $\lambda$ is zero at $\rho(\lambda)$, So
\begin{equation}
	5f\left(\frac{c}{T\lambda_m} \right) + \frac{c}{T\lambda_m}f'\left(\frac{c}{T\lambda_m} \right) = 0
\end{equation}
Let $x = \frac{c}{T\lambda_m}$ and rearranging terms we get
\begin{equation}
	\frac{df(x)}{f(x)} = -5\frac{dx}{x}
\end{equation}
The solution to this differential equation is 
\begin{equation}
	f(x)x^5 = constant \equiv k 
\end{equation}
\begin{equation}
	f(\frac{c}{T\lambda_m}) = k\frac{(\lambda _mT)^5}{c^5} \quad \text{Then,} \quad \rho(\lambda_m) = \frac{kT^5}{c}
\end{equation}
%%%%%%%%%%%%%%%%%
This is not physical solution, so only solution is $x = \frac{c}{T\lambda_m}$ is constant, i.e.
\begin{equation}
	\lambda_m T = \text{Const.}
	\label{w3}
\end{equation}
This is known as \textit{Wien's displacement law} or generally known as the \textit{Wien's law}.\rightHighlight{Wien was awarded the Nobel Prize for his work on the black-body radiation problem in 1911.} Any function which satisfies Wien's scaling law will obey Wien's displacement law. The first experimental verification of Wien's law was given by German physicists Otto Lummer, and Ernst Pringsheim in 1895 \cite{nauenberg_2016, lummer_LT_const}. \\

In 1896, Wien published a second paper on black-body radiation \cite{wien_1896}. He described a black-body spectrum using the Maxwell-Boltzmann distribution of atoms and assumed that the wavelength of radiation is the only function of its velocity (square of the velocity). He wrote energy density as
\begin{equation}
	\rho(\lambda, T) = F(\lambda)e^{-\frac{a}{\lambda T}}
\end{equation}
Now, using the Stefan-Boltzmann law, we have
\begin{equation}
	\int_0^\infty F(\lambda)e^{-\frac{a}{\lambda T}}d\lambda \propto T^4
	\label{w4}
\end{equation}
Wien defined $y := \frac{a}{\lambda T}$ and wrote expansion of function $F(\lambda)$ as 
\begin{align}
	F(\lambda) &= F\left(\frac{a}{yT}\right)  \nonumber\\
	&= c_0 + c_1\frac{yT}{a} + c_2\frac{y^2T^2}{a^2} + \dots + c_n\frac{y^nT^n}{a^n} + \dots \nonumber\\
	& c_{-1}\frac{a}{yT} + c_{-2}\frac{a^2}{y^2T^2} + \dots + c_{-n}\frac{a^n}{y^nT^n} + \dots
\end{align}
Putting into integral, we get
\begin{align}
	\int_0^\infty F(\lambda)e^{-\frac{a}{\lambda T}}d\lambda &= \frac{a}{T}\int_0^\infty F(\frac{a}{yT})e^{-y}\frac{dy}{y^2} \nonumber\\
	&= \sum_n c_n\frac{T^{n-1}}{a^{n-1}}\int_0^\infty e^{-y}y^{n-2}dy \nonumber\\
	&= \sum_n c_n\frac{T^{n-1}}{a^{n-1}}\Gamma(n-1)
\end{align}
This should be proportional to the $T^4$, so we must have $c_5$ is non-zero, and all others must be zero, which gives $F(\lambda) = \frac{C}{\lambda^5}$ where $C$ is constant. So energy density has a form
\begin{equation}
	\rho(\lambda, T) = \frac{C}{\lambda^5}e^{-\frac{a}{\lambda T}}\ .
\end{equation}
This is known as Wien's distribution law or Wien's approximation. This agrees with the equation (\ref{w2}) and gives a quite close distribution to experimental results.  In the same month, German physicist Friedrich Paschen found that his experimental data is best fitted to function \cite{paschen_1896}
\begin{equation}
	\rho(\lambda, T) = \frac{C_P}{\lambda^{5.56}}e^{-\frac{a_P}{\lambda T}}\ .
\end{equation}
Which is close to Wien's empirical distribution.
Wien did not give a rigorous physical argument for the empirical exponential factor. Many other attempts were made to find Kirchhoff's universal function, but this was the first closest function that was found. In 1899, Lummer and Pringsheim checked that the Wien's distribution agrees for small wavelengths, but at a larger wavelength and higher temperature, there is a systematic difference with experimental data.

\section{Enter Lord Rayleigh}
\begin{figure}[h]
	\caption{Lord Rayleigh \textit{(1842-1919) [Photo: Wikipedia Commons]}}
	\centering
	\includegraphics[width=4.2cm, height=5.5cm]{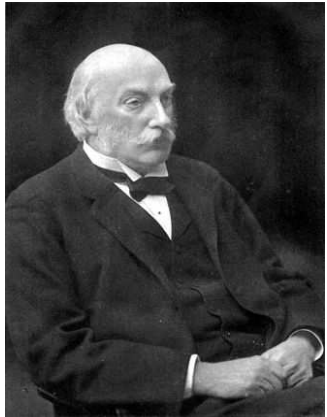}
	\label{Rayleigh}
\end{figure}

In June 1900, English mathematician and physicist John William Strutt, known as $3^{rd}$ Baron Rayleigh, gave another form of universal function in a paper ``Remarks upon the law of complete radiation". Rayleigh was the founder of the theory of sound, published many papers, and wrote two books on it. Using the analogy of the theory of sound, Rayleigh classically derived that the emission spectrum is proportional to the $T\lambda^{-4}$ \cite{rayleigh_1900}.  He assumed that radiation inside the cubical box formed a standing wave. His idea was to calculate the energy density according to 
\begin{equation}
	\rho(\nu, T) = [N(\nu)d\nu]\epsilon(\nu, T)
	\label{R1}
\end{equation}
where $N(\nu)d\nu$ is the number of standing waves per unit volume having a frequency in a range of $\nu$ to $\nu + d\nu$ and $\epsilon(\nu, T)$ is the mean energy of the wave. If $k$ is the wavenumber (reciprocal of $\lambda$), then the number of points for which $k$ lies between $k + dk$ is proportional to the $k^2dk$. According to the Maxwell-Boltzmann law, energy is equally divided among all modes. And since the energy of each mode is proportional to temperature T, the energy of radiation whose wavenumber is in the range $k$ and $k + dk$ will be proportional to the
\begin{equation}
	Tk^2dk\ .
\end{equation}
Or the energy of radiation whose wavelength is the range $\lambda$ and $\lambda + d\lambda$ will be
\begin{equation}
	\rho(\lambda, T)d\lambda \propto T\lambda^{-4}d\lambda
\end{equation}
In Rayleigh's first paper on radiation, he did not calculate the proportionality constant. The correct form of Rayleigh's law came five years later by Rayleigh himself and James Jeans. In between, Max Planck published two remarkable papers on radiation theory and introduced the idea of discretization of energy. In September 1900, Lummer and Pringsheim experimentally checked that Rayleigh's law agrees only for large wavelength. Lummer and Pringsheim put exponential factor $exp[-c_2/(\lambda T)^{\alpha}]$ and from the best-fitted data, they gave an empirical distribution function \cite{lummer_1899,lummer_1900}
\begin{equation}
	\rho(\lambda, T) = c_1T\lambda^{-4}\ exp[-c_2/(\lambda T)^{1.3}]
\end{equation}

\section{Enter Max Planck}
\begin{figure}[h]
	\caption{Max Planck \textit{(1858-1947) [Photo: Wikipedia Commons]}}
	\centering
	\includegraphics[width=4.2cm, height=5.5cm]{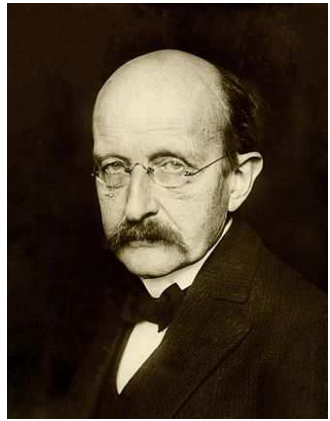}
	\label{Planck}
\end{figure}
The German physicist Max Karl Ludwing Planck was trying to justify Wien's distribution function using Heinrich Hertz's idea of electric dipole radiation. Form Kirchhoff's law, Planck understood that the black-body radiation is independent of the material of the cavity, so he presumed that the cavity walls are made of a collection of damped oscillators of electric dipoles (Hertzian resonators).\\

Planck considered the steady state case where the energy emitted by the accelerating charges via electromagnetic radiation should be equal to the energy absorbed by the oscillators. So oscillations are damped-driven. In 1899, he proved that the average energy of oscillator $U(\nu, T)$ is related to the energy density of radiation $\rho(\nu, T)$ in the following way \footnote{For further details and proof, readers can refer to Max Jammer book \textit{The Conceptual Development of Quantum Mechanics, Appendix A, pp. 470} and D. Ter. Haar book \cite{Haar_Book} \textit{Chepter 1, pp. 8}}
\begin{equation}
	\rho(\nu, T) = \frac{8\pi\nu^2}{c^3}U(\nu, T)
	\label{MP1}
\end{equation}

%%%%%%
In October 1900, Planck published the paper ``On the theory of the energy distribution law of the normal spectrum", where he guessed the function of the second derivative of entropy with respect to average energy and gave the first appearance of today known as Planck's distribution function. Planck made first simplest guess that 
\begin{equation}
	\dv[2]{S}{U} = - \frac{\alpha}{U}
	\label{MP5.0}
\end{equation}
Integrating the above equation with respect to $U$, we get
\begin{equation}
	\dv{S}{U} = -\alpha\ lnU + C
	\label{MP2}
\end{equation}
At constant volume, the fundamental thermodynamic equation $dU = TdS - PdV$ becomes
\begin{equation}
	\dv{S}{U} = \frac{1}{T}
	\label{MP3}
\end{equation}
Now, using the equations (\ref{MP2}) and (\ref{MP3}), we get
\begin{equation}
	U = A\ exp\left(-\frac{1}{\alpha T} \right)
\end{equation}
where $A$ and $\alpha$ depend upon the frequency of radiation. We get Wien's distribution function by putting this into equation (\ref{MP1}). \\
Now, if we say energy is linearly proportional to the temperature, then using equation (\ref{MP3}), we get
\begin{equation}
	\dv{S}{U} = \frac{\alpha'}{U}
\end{equation}
\begin{equation}
	\text{So,} \quad \dv[2]{S}{U} = -\frac{\alpha'}{U^2}
	\label{MP5}
\end{equation}
Planck then combined both equations (\ref{MP5.0}) and (\ref{MP5}) and made an intelligent guess that
\begin{equation}
	\dv[2]{S}{U} = -\frac{\alpha}{U(U+\beta)}
\end{equation}
Integrating the above equation with respect to $U$, we get
\begin{equation}
	\dv{S}{V} = \frac{\alpha}{\beta}\ ln\left(1 + \frac{\beta}{U} \right) + \text{Constant}
\end{equation}
Using equation (\ref{MP3}) and rearranging terms, we get
\begin{equation}
	U = \frac{a}{exp(b/T) -1}
\end{equation}
Putting into equation (\ref{MP1}) and comparing with Wien's scaling law (equation (\ref{w2.0})), we get the energy distribution function \cite{planck_origin_of_QM, planck_radiation_book, klein_1961, dougal_1976}
\begin{equation}
	\rho(\nu, T) = \frac{c_1\nu^3}{exp(c_2\nu/T) -1}\ .\\
\end{equation}

After a week, physicists Heinrich Rubens and Friedrich Kurlbaum published experimental verification of Planck's energy distribution for temperatures $85\ K$ to $1773\ K$, and it matched with all ranges of wavelengths \cite{dougal_1976}.  In December 1900, Planck published a second paper with a detailed explanation of the energy distribution where he introduced an idea of energy quantization and a universal constant known as Planck's constant \cite{Haar_Book, klein_1961}.\\

In the second paper, Planck thought that the total entropy of Hertzian resonators should be maximum at equilibrium. So it can be calculated from Boltzmann combinatorial method (microcanonical ensemble in statistical mechanics) \cite{planck_1901}.  Let $U$ be the average energy and $S$ be the entropy of a single oscillator. Then total energy of $N$ oscillators is $U_N = NU$, and total entropy is $S_N = NS$. Let $W$ be the total number of micro-states, then total entropy will be $S_N = k_B\ ln W$. Now, Planck thought that to calculate $W$ of $N$ oscillators having total energy $U_N$, we should take $U_N$ to be composed of discrete values of an element $\epsilon$ rather than a continuous one. So one can assume
\begin{equation}
	U_N = n\cdot\epsilon
	\label{Pq}
\end{equation}
where $n$ is a large integer. \rightHighlight{We can imagine this as a number of ways of putting $n$ identical objects into $N$ distinguishable box, where an empty box is allowed.}  Now, to calculate $W$, Planck calculated the number of ways to distribute $n$ bundles of energy $\epsilon$ among $N$ oscillators. Then,
\begin{equation}
	W = \frac{(N+n-1)!}{(N-1)!\ n!} \approx \frac{(N+n)!}{N!\ n!}
\end{equation}
Using the Stirling approximation for large number $N! \approx N^N$, we get
\begin{equation}
	W \approx \frac{(N+n)^{N+n}}{N^N\ n^n}
\end{equation}
Then, the total entropy will be
\begin{equation}
	S_N = k_B\left[(N+n)\ ln(N+n) - N\ lnN -n\ ln\ n\right]
\end{equation}
Now, using the $U_N = n\epsilon = NU$ and $S_N = NS$ we get
\begin{equation}
	S = k_B \left[ \left(1 + \frac{U}{\epsilon} \right)\ ln\left(1 + \frac{U}{\epsilon} \right) - \frac{U}{\epsilon}\ ln\frac{U}{\epsilon} \right]
	\label{P1}
\end{equation}
Using the $\rho = \frac{8\pi \nu^2}{c^3}U$ and Wien's scaling law, we get
\begin{equation}
	U = \nu f\left(\frac{T}{\nu} \right)
\end{equation}
Rearranging terms, we can write a different form
\begin{equation}
	T = \nu\ g\left(\frac{U}{\nu} \right)
\end{equation}
using the $\frac{1}{T} = \dv{S}{U}$, we get
\begin{equation}
	\pdv{S}{U} = \frac{1}{\nu}\ g\left(\frac{U}{\nu} \right)
\end{equation}
Integrating, we get
\begin{equation}
	S = g\left(\frac{U}{\nu} \right)
	\label{P2}
\end{equation}
Comparing equation (\ref{P1}) and (\ref{P2}), we must have $\epsilon \propto \nu$. The proportionality constant $``h"$ is known as Planck's universal constant. So putting $\epsilon = h\nu$ into equation (\ref{P1}) we get entropy
\begin{equation}
	S = k_B \left[ \left(1 + \frac{U}{h\nu} \right)\ ln\left(1 + \frac{U}{h\nu} \right) - \frac{U}{h\nu}\ ln\frac{U}{h\nu} \right]
\end{equation}
Again, using the  $\frac{1}{T} = \dv{S}{U}$, we get
\begin{equation}
	\frac{1}{T} = \frac{k_B}{h\nu}\ ln\left(1 + \frac{h\nu}{U} \right) 
\end{equation}
\begin{equation}
	\text{So,} \quad U = \frac{h\nu}{exp(h\nu/k_BT) - 1}
\end{equation}
So we get the energy distribution 
\begin{equation}
	\rho(\nu, T) = \frac{8\pi \nu^2}{c^3}\frac{h\nu}{exp(h\nu/k_BT) - 1}
\end{equation}
This is known as Planck's distribution function.\leftHighlight{Planck was awarded the Nobel Prize for his work on the quantum theory in 1919.} The universality of the Boltzmann constant was also given by Planck. Using experimental results of F. Kurlbaum on total radiation energy and Lummer and Pringsheim on Wien's displacement law Planck also calculated the value of the $k_B$ and $h$ \cite{planck_1901}.  The word "quanta" was first used by Planck for energy elements.  This strange non-classical theory of discretization of energy was uncomfortable to many scientists, including Planck himself. Albert Einstein's concern was that equation (\ref{MP1}) which Planck derived using the classical theory of absorption and emission of the energy by the oscillator, and in equation (\ref{Pq}), Planck defined the mean energy of the oscillator in a non-classical way. Scientists started taking Planck's idea seriously when Einstein explained Heinrich Hertz's experiment photoelectric effect by assuming that the energy of the light comes in quanta (light quanta) in 1905-06. \\

Einstein's concern about Planck's theory got resolved when Peter Debye \cite{jammer1966conceptual, born1989atomic} gave an alternative approach to Planck's theory using the idea from Rayleigh-Jeans theory. Debye considered electromagnetic radiation in a cubical cavity similar to the proper vibration of the crystal. So the number of vibrations in the interval $\nu$ to $\nu + d\nu$ is the same as in the calculation of Rayleigh and Jeans. The energy of the proper vibration of frequency $\nu$ can be from possible values $nh\nu$, where n is an integer. Using the Boltzmann probability distribution $P(n) = Aexp(-nh\nu/k_BT)$, the mean energy of the proper vibration will be
\begin{align}
	\epsilon(\nu, T) & = \frac{\sum_n nh\nu P(n)}{\sum_n P(n)} \\
	& = h\nu x\frac{1+ 2x + 3x^2 + \dots}{1+ x + x^2 + \dots}, \quad \text{where} \quad x = e^{-h\nu/k_BT} \\
	& = h\nu x\frac{1/(1-x)^2}{1/(1-x)} \\
	& = \frac{h\nu}{exp(h\nu/k_BT) -1}
\end{align}
Using the equation (\ref{R1}) and (\ref{J1}) gives the energy density 
\begin{equation}
	\rho(\nu, T) = \frac{8\pi \nu^2}{c^3} \frac{h\nu}{exp(h\nu/k_BT) -1}\\
\end{equation}

\section{Enter James Jeans}
\begin{figure}[h]
	\caption{James Jeans \textit{(1877-1946) [Photo: Wikipedia Commons]}}
	\centering
	\includegraphics[width=4.2cm, height=5.5cm]{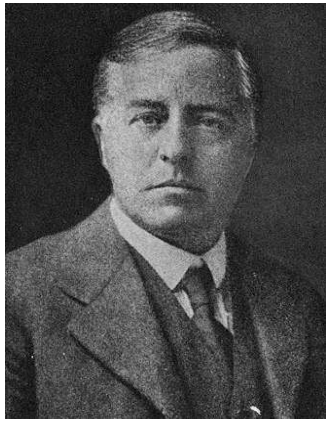}
	\label{Jeans}
\end{figure}
Even though Planck's distribution agreed with experimental data for many years, physicists could not believe the idea of quantization of energy. Rayleigh was still working on his idea of radiation in a cubical enclosure. In April 1905, he calculated a numerical constant in his previously proposed radiation energy distribution \cite{rayleigh_1905}. Rayleigh considered the standing wave $f(x,y,z,t) = Asin(k_xx+k_yy+k_zz)cos\omega t$ in cubical enclosure of length $l$. In order to satisfy the boundary condition ($f=0$ at $x,y$, or $z$ $= 0$ or $x,y$, or $z$ $= l$) we must have $k_x = \frac{n_x\pi}{l}$, $k_y = \frac{n_y\pi}{l}$ and $k_z = \frac{n_z\pi}{l}$ where $n_x, n_y, n_z \in \mathrm{N}$. Since $k^2 = k_x^2 + k_y^2 + k_z^2$ we have
\begin{equation}
	n_x^2 + n_y^2 + n_z^2 = \frac{k^2l^2}{\pi^2}
\end{equation}
which is a sphere of radius $\frac{kl}{\pi}$\\

Now, to calculate the total number of modes, we need to take eighth part of the volume of a sphere since $n_x, n_y,$ and $n_z$ can only take positive values. Rayleigh considered the positive as well as negative values. This mistake was pointed out by English physicist sir James Hopwood Jeans in May 1905 \cite{jeans_1905}.\\

With the Jeans correction factor, the total number of modes from $0$ to $k$ is
\begin{equation}
	\frac{1}{8}\frac{4}{3}\pi\left(\frac{kl}{\pi} \right)^3
\end{equation}
then the number of modes per unit volume in frequency range $\nu$ to $\nu + d\nu$ is
\begin{equation}
	N(\nu)d\nu = 2\frac{4\pi\nu^2}{c^3}d\nu 
	\label{J1}
\end{equation}
where we have used $k = \frac{2\pi\nu}{c}$ and factor `$2$' is multiplied since there are two polarization states of vibration. In terms of wavelength, the number of modes in the range $\lambda$ to $\lambda + d\lambda$ is
\begin{equation}
	N(\lambda)d\lambda = \frac{8\pi}{\lambda^4}d\lambda
\end{equation}
Now, since each mode is vibrating with average energy $k_BT$, from equation (\ref{R1}), we have the energy density 
\begin{equation}
	\rho(\lambda, T) = \frac{8\pi}{\lambda^4}k_BT
\end{equation}
This is known as `Rayleigh-Jeans law'. It agrees with experiments for large wavelength, but as we go towards a small wavelength (in the UV region of the black-body spectrum), energy density becomes very large. In 1911, Paul Ehrenfest called this the `Ultraviolet catastrophe'. But the problem was already solved by Planck. It took many years even for experts working in the filed to appreciate that.\\

%%place\leftHighlight or \rightHighlight depending on the margin (page) it appears
%%place exactly aligning to the text it has to appear
\section{Conclusion}
Textbooks usually do not explain Max Planck's original derivation of the black-body spectrum and how exactly he introduced the quantization of energy. Planck's first assumption that cavity walls can be considered as a collection of oscillators of electric dipoles came from Kirchhoff's law of universality. Instead of directly working on the energy distribution function like Rayleigh and Jeans, Planck used the microcanonical ensemble in statistical mechanics and tried to calculate the entropy of the oscillators. Planck showed that entropy is a function of the ratio of the average energy of oscillators and the frequency of radiation, and by comparing that with entropy calculated from the Boltzmann combinatorial method, he concluded that there should be some universal constant (Planck's constant $h$) such that energy of quanta is $h\nu$. Most of the textbook explains Wien's displacement law and Stefan-Boltzmann law as a consequence of Planck's distribution function, but historically Planck used Wien's scaling law to get the function of entropy of oscillator (equation (\ref{P2})). To derive the scaling law, Wien used the Stefan-Boltzmann $T^4$ law. So not only for a historical purpose but also to understand the black-body radiation rigorously, one needs to understand the original calculation and arguments of Kirchhoff, Stefan, Boltzmann, Wien, Rayleigh, Planck and Jeans in an actual timeline. \\

The logical way to explain the black-body radiation is to understand the Rayleigh-Jeans law first and appreciate that the black-body spectrum can not be explained using classical physics, and we need some different new formalism. Then by introducing the energy quantization and calculating the mean energy of Planck's cavity oscillators using the canonical ensemble in statistical mechanics, we can explain Planck's distribution function as a modification of Rayleigh-Jeans law.\\

Many physicists feel that Wien's and Planck's original derivations of their law are pretty long and not worth discussing, but we should at least appreciate the fact that without Kirchhoff's universal law, Stefan-Boltzmann $T^4$ law and Wien's scaling law, Planck may never have got the idea of energy quantization.

%%to place the figure caption in the margin

%%To place a figure 
%\begin{figure}[!b]
%\vskip 36pt
%\hskip-3.5cm %%to pull the figure in the margin area if it is a wider figure
% and placed in an even page
%\centering\includegraphics[width=5in, height=2.5in]{newton}
%\vskip 18pt
%\end{figure}

%%Use section* for unnumbered sections and \footnote for a normal footnote
\section*{Acknowledgement}
HM thanks DAE for the Disha fellowship and also thank his friend Debankit Priyadarshi (NISER, Bhubaneshwar) for giving important suggestions.

\section*{Appendix}
\subsection*{A. Derivation of the radiation pressure \label{A}}
Using the classical theory of electromagnetic radiation, we know that the momentum density is given by the ratio of the energy density and speed of light. \\
Let us say radiation is propagating, as shown in the figure.  Consider a tube of length $c\Delta t$ with a cross-section area of $L\times L$, making an angle $\theta$ with the normal to the walls. The energy contained in the tube is approximately 
\begin{equation}
	\rho_T\ c\Delta t\ L^2
\end{equation}
\begin{figure}[h]
	\caption{Imaginary tube of length $c\Delta t$ and cross-section area $L\times L$. The arrow shows the incident direction of radiation.}
	\centering
	\includegraphics[width=5cm, height=2.5cm]{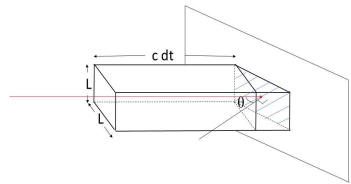}
	\label{Appendix_Fig}
\end{figure}
Then the energy per unit area of the wall is given by
\begin{equation}
	\rho_T\ c\Delta t\ \frac{L^2}{L \times \frac{L}{cos\theta}} = \rho_T\ c\Delta t cos\theta
\end{equation}
Thus, the momentum per unit area along the incident line is
\begin{equation}
	\rho_T \Delta t cos\theta
\end{equation}
But, the $cos\theta$ component of this momentum will contribute to the radiation pressure. And since the wall is perfectly reflecting, the net momentum transfer will be
\begin{equation}
	2\rho_T \Delta t cos^2\theta
\end{equation}
By dividing $\Delta t$, we get the force per unit area, i.e. pressure on the wall is 
\begin{equation}
	2\rho_T cos^2\theta
\end{equation}
The total pressure exerted can be calculated by taking the angular average. So,
\begin{equation}
	P = \frac{1}{4\pi}2\rho_T \int_0^{\pi/2}cos^2\theta sin\theta d\theta \int_0^{2\pi}d\phi
\end{equation}
The $\theta$ integration gives the factor $1/3$, So the radiation pressure is
\begin{equation}
	P = \frac{\rho_T}{3}
\end{equation}

%%References section
\bibliographystyle{plain} 
\bibliography{main}

\begin{thebibliography}{10}

\bibitem{bartoli_1884}
Adolfo Bartoli.
\newblock Il calorico raggiante e il secondo principio di termodinamica.
\newblock {\em Il Nuovo Cimento (1877-1894)}, 15(1):193--202, 1884.

\bibitem{born1989atomic}
Max Born, Roger~John Blin-Stoyle, and John~Michael Radcliffe.
\newblock {\em Atomic physics}.
\newblock Courier Corporation, 1989.

\bibitem{boya_2004}
Luis~J Boya.
\newblock The thermal radiation formula of planck (1900).
\newblock {\em arXiv preprint physics/0402064}, 2004.

\bibitem{chang_2002}
Hasok Chang.
\newblock Rumford and the reflection of radiant cold: Historical reflections
  and metaphysical reflexes.
\newblock {\em Physics in Perspective}, 4(2):127--169, 2002.

\bibitem{darrigol_1988}
Olivier Darrigol.
\newblock Statistics and combinatorics in early quantum theory.
\newblock {\em Historical studies in the physical and biological sciences},
  19(1):17--80, 1988.

\bibitem{dougal_1976}
RC~Dougal.
\newblock The presentation of the planck radiation formula (tutorial).
\newblock {\em Physics Education}, 11(6):438, 1976.

\bibitem{draper_1847}
John~William Draper.
\newblock On the production of light by heat.
\newblock {\em Journal of the Franklin Institute}, 44(2):122--128, 1847.

\bibitem{gearhart}
Clayton~A Gearhart.
\newblock Planck, the quantum, and the historians.
\newblock {\em Physics in perspective}, 4(2):170--215, 2002.

\bibitem{jammer1966conceptual}
Max Jammer.
\newblock {\em The conceptual development of quantum mechanics}.
\newblock McGraw-Hill, 1966.

\bibitem{jeans_1905}
JH~Jeans.
\newblock The dynamical theory of gases and of radiation.
\newblock {\em Nature}, 72(1857):101--102, 1905.

\bibitem{kelly_1965}
F~Kelly.
\newblock On kirchhoff's law and its generalized application to absorption and
  emission by cavities.
\newblock In {\em 2nd Aerospace Sciences Meeting}, page 135, 1965.

\bibitem{kirchhoff_1859}
Gustav Kirchhoff.
\newblock Uber den zusammenhang zwischen emission und absorption von licht und.
  w{\"a}rme.
\newblock {\em Monatsberichte der Akademie der Wissenschaften zu Berlin}, pages
  783--787, 1859.

\bibitem{Kirchhoff_1860}
Gustav Kirchhoff.
\newblock On the relation between the radiating and absorbing powers of
  different bodies for light and heat.
\newblock {\em The London, Edinburgh, and Dublin Philosophical Magazine and
  Journal of Science}, 20(130):1--21, 1860.

\bibitem{klein_1961}
Martin~J Klein.
\newblock Max planck and the beginnings of the quantum theory.
\newblock {\em Archive for History of Exact Sciences}, 1(5):459--479, 1961.

\bibitem{kuhn_1987}
Thomas~S Kuhn.
\newblock {\em Black-body theory and the quantum discontinuity, 1894-1912}.
\newblock University of Chicago Press, 1987.

\bibitem{lummer_1899}
Otto Lummer and Ernst Pringsheim.
\newblock Die vertheilung der energie im spectrum des schwarzen k{\"o}rpers.
\newblock {\em Verhandlungen der Deutsche Physikalische Gesellschaft},
  1(23):215, 1899.

\bibitem{lummer_1900}
Otto Lummer and Ernst Pringsheim.
\newblock {\em Uber die Strahlung des schwarzen Korpers fur lange Wellen}.
\newblock Barth, 1900.

\bibitem{lummer_LT_const}
Otto Lummer and Ernst Pringsheim.
\newblock Kritisches zur schwarzen strahlung.
\newblock {\em Annalen der Physik}, 311(9):192--210, 1901.

\bibitem{masalov_2019}
Anatoly~V Masalov.
\newblock First experiments on measuring light pressure i (pyotr nikolaevich
  lebedev).
\newblock In {\em Quantum Photonics: Pioneering Advances and Emerging
  Applications}, pages 425--453. Springer, 2019.

\bibitem{Maxwell_Book}
James~Clerk Maxwell and Peter Pesic.
\newblock {\em Theory of heat}.
\newblock Courier Corporation, 2001.

\bibitem{nauenberg_2016}
Michael Nauenberg.
\newblock Max planck and the birth of the quantum hypothesis.
\newblock {\em American Journal of Physics}, 84(9):709--720, 2016.

\bibitem{needell_980}
Allan~A Needell.
\newblock {\em Irreversibility and the failure of classical dynamics: Max
  Planck's work on the quantum theory 1900-1915}.
\newblock Yale University, 1980.

\bibitem{Leopoldo_1831}
Leopoldo Nobili and Macedonio Melloni.
\newblock Recherches sur plusieurs ph{\'e}nom{\`e}nes calorifiques entreprises
  au moyen du thermo-multiplicateur.
\newblock In {\em Annales de Chimie et de Physique}, volume~48, pages 198--218,
  1831.

\bibitem{paschen_1896}
F~Paschen.
\newblock {\"U}ber gesetzm{\"a}ssigkeiten in den spectren fester k{\"o}rper.
\newblock {\em Annalen der Physik}, 294(7):455--492, 1896.

\bibitem{planck_1901}
Max Planck.
\newblock On the law of the energy distribution in the normal spectrum.
\newblock {\em Ann. Phys}, 4(553):1--11, 1901.

\bibitem{planck_origin_of_QM}
Max Planck.
\newblock {\em The origin and development of the quantum theory}.
\newblock Clarendon Press, 1922.

\bibitem{planck_radiation_book}
Max Planck and Niels Bohr.
\newblock {\em Quantum Theory}.
\newblock Flame Tree Publishing., 2019.

\bibitem{rayleigh_1889}
Lord Rayleigh.
\newblock Liii. on the character of the complete radiation at a given
  temperature.
\newblock {\em The London, Edinburgh, and Dublin Philosophical Magazine and
  Journal of Science}, 27(169):460--469, 1889.

\bibitem{rayleigh_1900}
Lord Rayleigh.
\newblock Remarks upon the law of complete radiation.
\newblock {\em Philosophical Magazine}, 49:539--540, 1900.

\bibitem{rayleigh_1905}
Lord Rayleigh.
\newblock The dynamical theory of gases and radiation.
\newblock {\em Nature}, 72:54--55, 1905.

\bibitem{robitaille_2008}
Pierre-Marie Robitaille.
\newblock Blackbody radiation and the carbon particle.
\newblock {\em Progress in Physics}, 3:36, 2008.

\bibitem{schirrmacher_2003}
Arne Schirrmacher.
\newblock Experimenting theory: The proofs of kirchhoff's radiation law before
  and after planck.
\newblock {\em Historical studies in the physical and biological sciences},
  33(2):299--335, 2003.

\bibitem{siegel_1976}
Daniel~M Siegel.
\newblock Balfour stewart and gustav robert kirchhoff: Two independent
  approaches to" kirchhoff's radiation law".
\newblock {\em Isis}, 67(4):565--600, 1976.

\bibitem{stefan_1879}
J~Stefan.
\newblock Uber die beziehung zwischen der warmestrahlung und der temperatur,
  sitzungsberichte der mathematisch-naturwissenschaftlichen classe der
  kaiserlichen.
\newblock {\em Akademie der Wissenschaften}, 79:S--391, 1879.

\bibitem{stewart_1858}
Balfour Stewart.
\newblock I.—an account of some experiments on radiant heat, involving an
  extension of prevost's theory of exchanges.
\newblock {\em Earth and Environmental Science Transactions of The Royal
  Society of Edinburgh}, 22(1):1--20, 1858.

\bibitem{Haar_Book}
Dirk Ter~Haar.
\newblock {\em The Old Quantum Theory: The Commonwealth and International
  Library: Selected Readings in Physics}.
\newblock Elsevier, 2016.

\bibitem{tyndall_1865}
J~Tyndall.
\newblock Heat considered as a mode of motion, london, 1865.

\bibitem{wien_1896}
Willy Wien.
\newblock Xxx. on the division of energy in the emission-spectrum of a black
  body.
\newblock {\em The London, Edinburgh, and Dublin Philosophical Magazine and
  Journal of Science}, 43(262):214--220, 1896.

\end{thebibliography}

\end{document}